\documentclass[aps,pra,superscriptaddress,one-column]{revtex4-2}

\usepackage[utf8]{inputenc}
\usepackage[version=3]{mhchem}
\usepackage{amsmath}
\usepackage{amsfonts}
\usepackage{braket}
\usepackage[draft]{graphicx}
\usepackage{color}
\usepackage{float}
\usepackage{xfrac}

\makeatletter
\renewcommand\frontmatter@abstractwidth{\dimexpr\textwidth-1in\relax}
\makeatother

\DeclareMathOperator\erf{erf}

\begin{document}

\title{Hybrid basis and multi-center grid method for strong-field processes}

\author{Kyle A. Hamer}
\email{kyle-hamer@ucf.edu}
\affiliation{Department of Physics, University of Central Florida, Orlando, FL, 32816, USA}

\author{Heman Gharibnejad}
\affiliation{Computational Physics Inc., Springfield, VA, 22151, USA}

\author{Luca Argenti}
\affiliation{Department of Physics, University of Central Florida, Orlando, FL, 32816, USA}

\author{Nicolas Douguet}
\email{nicolas.douguet@ucf.edu}
\affiliation{Department of Physics, University of Central Florida, Orlando, FL, 32816, USA}

\begin{abstract}
We present a time-dependent framework that combines a hybrid Gaussian–FEDVR basis with a multicenter grid to simulate strong-field and attosecond dynamics in atoms and molecules. The method incorporates the construction of the orthonormal hybrid basis, the evaluation of electronic integrals, a unitary time-propagation scheme, and the extraction of optical and photoelectron observables. Its accuracy and robustness are benchmarked on one-electron systems such as atomic hydrogen and the dihydrogen cation (H$_2^+$) through comparisons with essentially-exact reference results for bound-state energies, high-harmonic generation spectra, photoionization cross sections, and photoelectron momentum distributions. This work establishes the groundwork for its integration with quantum-chemistry methods, which is already operational but will be detailed in future work, thereby enabling {\it ab initio} simulations of correlated polyatomic systems in intense ultrafast laser fields.
\end{abstract}

\maketitle

\section{Introduction}

Advances in laser technology over the past three decades have enabled the production of controllable, intense, few-cycle optical pulses \cite{sartania1997, brabec2000, seidel2018, furch2022}, allowing us to break the ``femtosecond barrier'' and ushering in the era of attosecond science \cite{corkum2007, plaja2013, nisoli2017, biegert2021, cruz-rodriguez2024}. These ultrashort pulses of light promise to provide a better understanding of chemical processes by allowing us to induce, probe, and perhaps steer electron dynamics on its natural timescale. In recent years, attosecond science has made possible the reconstruction of photoelectron wavepackets \cite{klunder2013, bello2018, fuchs2021}, development of efficient techniques for chirality discrimination \cite{beaulieu2017, bloch2021, fehre2021, ayuso2022}, induction of charge migration in molecules \cite{calegari2014, kraus2015, lara-astiaso2018, grell2023, he2023}, among many other applications \cite{sansone2010, sainadh2019, tuthill2022, chini2014, haas2016}. 

These remarkable experimental achievements present theorists with new and unique challenges because of the need to address the complex dynamics of many electrons within a high-intensity, time-varying optical field. As a result, there are several advanced theoretical methods (with varying degrees of computational efficiency and accuracy) to support, predict, and interpret the strong-field and ultrafast phenomena explored in laboratories \cite{oliveira2015, armstrong2021}. These include: the time-dependent configuration interaction (TD-CI) method \cite{rohringer2006, greenman2010, hochstuhl2012, wozniak2022}, the time-dependent complete-active-space self-consistent-field (TD-CASSCF) method \cite{sato2013, miyagi2013, nest2008}, time-dependent density-functional theory (TD-DFT) \cite{marques2004, mauger2025}, and the R-matrix with time dependence (RMT) \cite{brown2020, bondy2024} method, among others.

In this paper, we present recent developments around ATTOMESA (ATTOsecond Molecular Electronic Structure Application), a new time-dependent quantum chemistry code designed for attosecond science, with a particular emphasis on the interaction of atoms and molecules with intense laser fields. The general structure of ATTOMESA shares key features with other hybrid approaches such as the time-dependent restricted-active-space configuration-interaction (TD-RASCI) method \cite{hochstuhl2012}, XCHEM \cite{marante2014, marante2017}, and ASTRA \cite{randazzo2023}. Specifically, it combines a quantum-chemical description using Gaussian-type orbitals (GTOs) near the atomic or molecular core with a finite-element discrete-variable representation (FEDVR) to describe the photoelectron dynamics at larger distances. This hybrid approach enables the \textit{ab initio} treatment of atomic and molecular processes in ultrashort pulses, such as core-hole spectroscopy, as well as processes in intense fields, including high-harmonic generation (HHG) and strong-field ionization..

While ATTOMESA has already been applied to strong-field phenomena in two-electron systems, such as HHG in helium \cite{bondy2024} and H$_2$, the present work focuses on the one-electron formulation to validate the method and demonstrate its robustness by benchmarking against essentially-exact results. We provide a detailed account of the hybrid GTO–FEDVR basis, the hybrid numerical quadrature used to evaluate electronic integrals, the orthonormalization procedure, the propagation of the time-dependent Schr\"{o}dinger equation, and the construction of physical observables relevant to strong-field physics. A complete presentation of the multi-electron implementation, including its quantum chemistry components, will be provided in a future contribution.

The paper is organized as follows. Section 2 introduces the one-electron formalism implemented in ATTOMESA and outlines the calculation of observables relevant to attosecond science. In Section 3, we benchmark the method using atomic hydrogen, a one-electron model potential in helium, and the dihydrogen cation, focusing on high-harmonic generation and photoelectron spectroscopy. Section 4 summarizes the strengths of ATTOMESA and outlines future directions for the development and application of its multi-electron capabilities.

Unless otherwise specified, all quantities are expressed in atomic units throughout this manuscript.

\section{Theoretical Approach}

Section 2.1 introduces the domain decomposition of ATTOMESA’s simulation volume into a molecular region $\Omega_{m}$ and an external region $\Omega_{e}$. Section 2.2 reviews the finite-element discrete-variable representation (FEDVR). Section 2.3 presents the hybrid basis built from Gaussian-type orbitals (GTOs) and FEDVR functions. Section 2.4 describes the evaluation of electronic integrals. Section 2.5 constructs an orthonormal hybrid basis using high-accuracy quadrature in $\Omega_{m}$, correcting the fact that the FEDVR functions are initially approximately orthonormal there under the accurate quadrature. Section 2.6 details the time-propagation scheme. Finally, Section 2.7 defines the physical observables computed by ATTOMESA.

\subsection{Partitioning of Physical Space}

\begin{figure}[htbp]
    \centering
    \includegraphics[width=0.95\columnwidth]{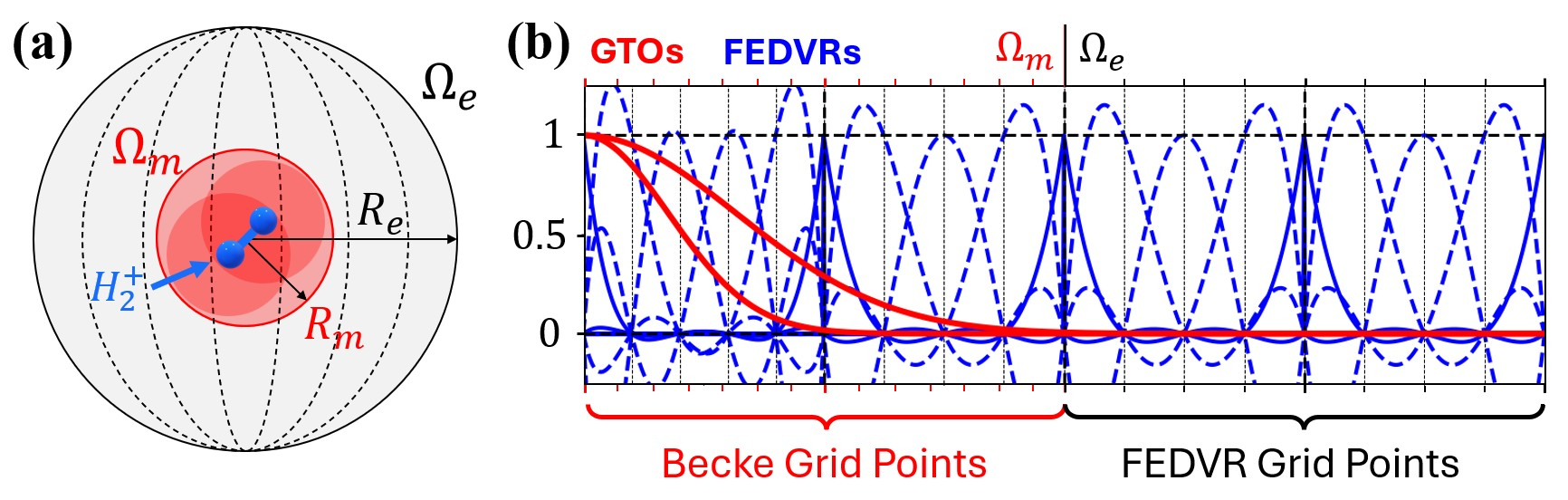}
    \caption{(a) ATTOMESA's simulation volume, with a dihydrogen cation in the center; (b) prototypical radial basis functions used in ATTOMESA's hybrid basis: the Gaussian-type orbitals (red) reside only within the molecular region $\Omega_{m}$, while the FEDVRs (blue) reside within both $\Omega_{m}$ and the external region $\Omega_{e}$. For clarity, here we assume that there is only one atom located at the center of the spherical volume and so only one set of Gaussian-type orbitals centered at the origin. The tick marks at the bottom and top of the plot denote the radial grid points used in the computation of the electronic integrals.}
    \label{fig:1}
\end{figure}

The spatial partition used in ATTOMESA is illustrated in Fig.~\ref{fig:1}(a), where the atom or molecule of interest is positioned at the center of a spherical volume, which is separated into a molecular region of radius $R_m$, denoted $\Omega_{m}$, and an external region of radius $R_e$, denoted $\Omega_{e}$.

ATTOMESA employs a hybrid basis \cite{rescigno2005, yip2008} in which atomic or molecular orbitals, constructed from linear combinations of Gaussian-type orbitals (GTOs) \cite{boys1960, schaefer1977} centered on each atom, are complemented by finite-element discrete-variable representation (FEDVR) functions \cite{rescigno2000, power2012} centered at the origin. These molecular orbitals (MOs) are negligible outside the molecular region $\Omega_{m}$, while the FEDVR functions span both the molecular region $\Omega_{m}$ and the external region $\Omega_{e}$. The electronic integrals in the molecular region are computed on a multicenter grid based on a modified version of Becke's scheme \cite{becke1988, gharibnejad2021}. At sufficiently large distances from the atomic centers, the multicenter grid transitions smoothly to a spherical form, enabling a seamless connection with the FEDVR grid in $\Omega_{e}$, where the advantageous properties of grid-based FEDVR functions can be fully exploited. This hybrid basis and quadrature method, illustrated in Fig.~\ref{fig:1}(b), provides an accurate and efficient description of both electronic bound states and strong-field photoelectron dynamics.

\subsection{Radial FEDVR Functions}

In the finite-element method (FEM), the radial part of the basis functions are represented by a set of compact polynomials, each defined within one of the $N$ elements. These elements are defined by a set of boundary points $0 \leq R_{1} < R_{2} < \cdots < R_{N+1}$. Within each element $i = 1, \ldots, N$, one can define discrete variable representation (DVR) basis functions in terms of Lagrange-Lobatto interpolating polynomials $f_{i,m}(r)$ \cite{manopoulos1988, rescigno2000} as 
\begin{align}\label{eq:proto_fedvrs}
    f_{i,m}(r) &= \prod_{j \neq m} \frac{r - r_{i,j}}{r_{i,m} - r_{i,j}}, 
    && R_{i} \leq r \leq R_{i+1}, \nonumber \\
    f_{i,m}(r) &= 0, && \text{otherwise},
\end{align}
such that $f_{i,m}(r_{i,m^\prime})=\delta_{m,m^\prime}$. The $n$ mesh points $r_{i,m}$, for $m = 1, \dots, n$, are obtained from a Gauss–Lobatto quadrature \cite{krylov2006, manopoulos1988}, where the endpoints are included as quadrature points ($r_{i,1} = R_{i}$ and $r_{i,n} = R_{i+1}$), and weights $w_{i,m}$ are associated with each point to approximate the integral
\begin{equation}
    \int_{R_{i}}^{R_{i+1}} p(r)\, \mathrm{d}r \approx \sum_{m=1}^{n} p(r_{i,m})\, w_{i,m}
    \label{eq:quadrature}
\end{equation}
for a given radial function $p(r)$. Equation \eqref{eq:quadrature} is exact when $p(r)$ is a polynomial of degree $\le 2n-3$. Note that the number of points $n$ within each element may differ.

Prototypical DVR functions of Eq.~(\ref{eq:proto_fedvrs}) are shown in Fig.~\ref{fig:1}(b), in blue. In this example, there are four elements in total, the first two belonging to $\Omega_{m}$ and the second two belonging to $\Omega_{e}$. At each of the points defined by the Gauss-Lobatto quadrature within each element (vertical black dashed lines), only one of the basis functions has a value of 1 while the rest are 0. In the context of the Gauss-Lobatto quadrature, these basis functions are therefore orthogonal:
\begin{align}
    \int_{0}^{\infty} f_{i,m}(r)\, f_{i^{\prime},m^{\prime}}(r)\, \mathrm{d}r &= \delta_{i,i^{\prime}} \int_{R_{i}}^{R_{i+1}} f_{i,m}(r)\, f_{i^{\prime},m^{\prime}}(r)\, \mathrm{d}r \nonumber\\
    &= \delta_{i,i^{\prime}} \sum_{j=1}^{n} f_{i,m}(r_{i,j})\, f_{i,m^{\prime}}(r_{i,j})\, w_{i,j} = \delta_{i,i^{\prime}}\, \delta_{m,m^{\prime}}\, w_{i,m}
\end{align}
The DVR functions have an underlying continuous representation which allows one to evaluate a wavefunction at any arbitrary point $r$ once the expansion coefficients are known.

The continuity condition between elements can be applied by combining the piecewise functions $f_{i,n}$ and $f_{i+1,1}$ at the boundary between two adjacent elements, into a single \emph{bridge} function $\chi_{i,1}$ (solid blue lines in Fig.~\ref{fig:1}(b)). Therefore, we can construct a continuous basis that spans all elements by defining 
\begin{equation}\label{eq:def_fedvrs}
    \begin{cases}\chi_{i,1}(r) = (f_{i,n}(r) + f_{i+1,1}(r))/\sqrt{w_{i,n} + w_{i+1,1}} \\ \chi_{i,m}(r) = f_{i,m}(r)/\sqrt{w_{i,m}}\ , \quad m = 2 \ldots (n-1), \end{cases}\;
\end{equation}
leading to an orthonormal basis under Gauss-Lobatto quadrature:
\begin{equation}
\int_0^\infty \chi_{i,m}(r)\chi_{i^{\prime},m^{\prime}}(r)dr\approx \sum_{j=1}^n \chi_{i,m}(r_{i,j})\chi_{i^{\prime},m^{\prime}}(r_{i^{\prime},j})w_{i,m}=\delta_{i,i^{\prime}}\delta_{m,m^{\prime}}.
\end{equation}
It also follows that the DVR functions give a diagonal representation of any local radial operator $V(r)$ since
\begin{equation}
\int_0^\infty \chi_{i,m}(r)V(r)\chi_{i^{\prime},m^{\prime}}(r)dr\approx \sum_{j=1}^n \chi_{i,m}(r_{i,j})V(r_{i,j})\chi_{i^{\prime},m^{\prime}}(r_{i^{\prime},j})w_{i,j}=V(r_{i,m})\delta_{i,i^{\prime}}\delta_{m,m^{\prime}}.
\end{equation}

\subsection{Hybrid Basis and Quadrature Approach}

The hybrid basis used in this work combines two families of functions. The first consists of orthonormal MOs, $\varphi_\sigma$, $\sigma=1,\dots,N_o$, constructed from \emph{contracted} Gaussian functions, following the standard procedure used in molecular quantum chemistry codes. Each contracted function is a linear combination of Cartesian-Gaussian primitives centered on each atom at  
$\vec R_{a} = (X_a, Y_a, Z_a)$,  
of the form  
\begin{equation}
G^a_{\Gamma}(\vec r) = N_\Gamma \,(x - X_a)^{l_\Gamma} (y - Y_a)^{m_\Gamma} (z - Z_a)^{n_\Gamma} \,
e^{-\alpha_\Gamma \, |\vec r - \vec R_a|^2},
\end{equation}  
where $\Gamma$ indexes the primitive basis functions, $l_\Gamma, m_\Gamma, n_\Gamma$ are integers which describe the symmetry of the GTO, $N_\Gamma$ is the normalization constant, and $\alpha_\Gamma$ is the Gaussian exponent. In multi-electron calculations, optimized molecular orbitals can be generated in various ways, such as from Hartree–Fock (HF) calculations, from multi-reference configuration state functions (MCSCF), or as natural orbitals obtained by diagonalizing the one-particle density matrix from a configuration interaction (CI) calculation. For the one-electron problems considered here, the MOs are simply obtained by diagonalizing the Hamiltonian in the GTO basis. 

The GTOs are complemented by spherical FEDVR functions of the form
\begin{equation}
\chi_{\beta}(\vec r)=\frac{1}{r}\chi_{i,m}(r)X_{\ell m_\ell}(\hat{r}),
\end{equation}
where $r=|\vec r|$, $X_{\ell m_\ell}(\hat r)$ are real spherical harmonics, and $\beta\equiv \{i,m,\ell,m_\ell\}$ collects the indices of the radial element $i$, the local DVR function $m$, and the angular quantum numbers $(\ell,m_\ell)$. The FEDVR functions are defined on the radial interval $[r_0,R_e]$, with $r_0=0$ in this work. Note that from this point on, ``FEDVR functions'' may be shortened to ``FEDVRs''.

Within the molecular region $\Omega_m$, we use a multi-center grid constructed from a modified version of the scheme originally introduced by Becke in 1988 \cite{becke1988}. As the method is described in detail in \cite{gharibnejad2021}, we only briefly summarize its key features. In Becke's scheme, space is divided into overlapping ``fuzzy'' Voronoi polyhedra \cite{becke1988}, where weight functions $w_a(\vec r)$ for each atomic center $a$ form a partition of unity $\left(\sum_{a} w_{a}(\vec{r}) = 1\ \forall\ \vec{r}\right)$, with the additional requirement that $w_a(\vec R_b)=\delta_{ab}$. The smooth weight functions are constructed as the product of switch-off functions between pairs of atoms, automatically fulfilling this criterion. Using Becke's partitioning, electronic integrals can be split into separate atomic components as
\begin{equation}
    I = \int_{\Omega_m} f(\vec{r})\, \mathrm{d}^{3}\vec{r} = \sum_{a} \int_{\Omega_{m}} w_{a}(\vec{r})\, f(\vec{r})\, \mathrm{d}^{3}\vec{r},
\end{equation}
whose argument is regular everywhere except at a single nucleus. Every atomic integration region is subdivided into a set of spherical shells, where 3D quadrature points are constructed as a product of Gauss–Legendre radial nodes and a Lebedev angular grid \cite{lebedev1975}.

In Becke’s original scheme \cite{becke1988}, each atomic grid must span the entire continuum quantization volume, greatly increasing the number of grid points and leading to severe limitation in the treatment of molecules with many centers. Moreover, this construction introduces a mismatch at the interface between the molecular and external spherical regions that can lead to a loss of accuracy. In the modified Becke scheme \cite{gharibnejad2021}, we confine the atomic grids by modifying the atomic weight as $w_a^{At}(\vec r)=w_a(\vec r)f^{At}(|\vec{r} - \vec{R}_{a}|)$, where $f^{At}$ is a monotonically decaying factor that smoothly switches off the weight associated with a shell of radius $R^{At}$ centered at atom $a$. While various functional forms are possible, we use the smooth, monotonic cutoff
\begin{equation}
f^{At}(r)=\frac{1}{2}\left\{1-\erf\left[\alpha\left(\frac{r}{R^{At}}-\frac{1}{2}\right)\right]\right\},    
\end{equation}
where $\erf$ denotes the error function. We chose $\alpha$ so that the function $f^{At}$ is effectively unity at the atomic center and negligible for $r\ge R^{At}$ to accuracy $\delta<10^{-15}$. We add a single-center “master” grid placed at the same molecular origin as the external grid, choosing element boundaries that coincide with those of the FEDVR functions. The master-grid weights are defined as $w_{0}(\vec{r}) = 1 - \sum_{a} w_{a}(\vec{r})$, with $w_{0}(\vec{R}_{a}) = 0\ \forall a$. The atomic grids lie well within the molecular region $\Omega_m$, while the interstitial and asymptotic parts of the integrand are captured by the master grid. The master grid has radius $R_m$, so that it carries essentially all the weight near the boundary between $\Omega_m$ and $\Omega_e$. Consequently, the same electronic integral can be written as
\begin{equation}
    I = \int_{\Omega_m} f(\vec{r})\, \mathrm{d}^{3}\vec{r} = \left(\sum_{a} \int_{\Omega_{m}} w^{At}_{a}(\vec{r})\, f(\vec{r})\, \mathrm{d}^{3}\vec{r}\right) + \int_{\Omega_{m}} w_{0}(\vec{r})\, f(\vec{r})\, \mathrm{d}^{3}\vec{r}\ .
\end{equation}
Using multiple atomic-shell partitions, we benchmarked the modified Becke scheme against various analytically known integrals and obtained results that are accurate to nearly machine precision. Within $\Omega_m$, the FEDVR functions are treated on the same footing as the GTOs, i.e., we do not employ the FEDVR quadrature rule but instead evaluate all integrals directly, yielding results that are essentially exact. In particular, quadrature rules of higher order than Gauss–Lobatto may be employed, provided the radial element partition remains consistent with the construction of the FEDVR functions, so that derivative discontinuities at element interfaces are handled consistently and high-order accuracy is preserved.

In the external region $\Omega_e$, electronic integrals are evaluated using Gauss–Lobatto quadrature, consistent with the construction of the FEDVR functions. In this representation, the quadrature rules render any local operator diagonal in the radial coordinate, enabling highly efficient evaluation of electronic integrals as well as time propagation outside the molecular region.

\subsection{Calculation of electronic integrals}

For the one-electron systems considered in this work, the relevant matrices are the field-free Hamiltonian $\hat{H}_0=\hat{T}+\hat{V}$ and the light-interaction Hamiltonian $\hat{H}_{{\rm int}}(t)=-\vec{\mu}\cdot\vec{E}(t)$, both expressed in the hybrid GTO-FEDVR basis described in the previous subsection. Here, $\hat{T}$ is the kinetic-energy operator, $\hat{V}$ is the electrostatic potential, and $\vec{\mu}=-\vec{r}$ is the dipole operator. In the molecular region, for a given operator $\hat{\mathcal{O}}$, three types of matrix elements must be evaluated: $\braket{\varphi_{\sigma} | \hat{\mathcal{O}} | \varphi_{\sigma^{\prime}}}$ (MO-MO integrals), $\braket{\varphi_{\sigma} | \hat{\mathcal{O}} | \chi_\beta}$ (MO–FEDVR integrals), and $\braket{\chi_\beta | \hat{\mathcal{O}} | \chi_{\beta^\prime}}$ (FEDVR-FEDVR integrals). All integrals are evaluated after tabulating the GTOs, the action of the Laplacian on the GTOs, and the FEDVR functions on Becke’s grid. The only exception is the kinetic energy operator between two FEDVRs, which can be computed exactly in Gauss-Lobatto quadrature \cite{rescigno2000}, as
\begin{equation}
\langle\chi_\beta|\hat{T}|\chi_{\beta^\prime}\rangle=\frac{1}{2}(\delta_{i,i^\prime}+\delta_{i,i^\prime\pm 1})\int_0^\infty dr \frac{d}{dr}\chi_{i,m}(r)\frac{d}{dr}\chi_{i^\prime,m^\prime}(r)+\delta_{ii^\prime}\delta_{mm^\prime}\delta_{\ell\ell^\prime}\delta_{m_\ell m_{\ell}^\prime}\frac{\ell(\ell+1)}{2r_{i,m}^2}.
\end{equation}

In the external region $\Omega_{e}$, the Gauss-Lobatto quadrature is the natural choice for the evaluation of electronic integrals, consistent with the construction of the FEDVRs. In this context, any local operator, such as the potential energy operator $\hat{V}$, has a diagonal representation in the radial coordinate:
\begin{align}
    \braket{\chi_{\beta} | \hat{V} | \chi_{\beta^\prime}} &= \iiint_{\mathbb{R}^{3}} \big{(}\chi_{i,m}(r)\, X_{\ell}^{m_\ell}(\Omega)\big{)}\, V(r, \Omega)\, \big{(}\chi_{i^{\prime}, m^{\prime}}(r)\, X_{\ell^\prime}^{m_\ell^\prime}(\Omega)\big{)}\, \mathrm{d}r\, \mathrm{d}\Omega \nonumber\\
    &= \delta_{ii^{\prime}}\, \delta_{mm^{\prime}} \iint_{\mathbb{S}^{2}} X_{\ell}^{m_\ell}(\Omega)\, V(r
    _{i,m}, \Omega)\, X_{\ell^\prime}^{m_\ell^\prime}(\Omega)\, \mathrm{d}\Omega \nonumber\\
    &= \delta_{ii^{\prime}}\, \delta_{mm^{\prime}} \sum_{\kappa} w_{\kappa}\, X_{\ell}^{m_\ell}(\Omega_{\kappa})\, V(r_{i,m}, \Omega_{\kappa})\, X_{\ell^\prime}^{m_\ell^\prime}(\Omega_{\kappa}).
\end{align}
where $\Omega_{\kappa}$ and $w_{\kappa}$ are the points and weights, respectively, defined by the Lebedev quadrature used for the angular integration. 

The only exception arises for the bridge functions spanning the $\Omega_m$ and $\Omega_e$ regions (called the connectors $C$), whose matrix elements are evaluated by splitting the electronic integral into two parts. The part that lies in the molecular region $\Omega_{m}$ is computed using the Becke quadrature, while the part that lies within the external region $\Omega_{e}$ is computed using the FEDVR quadrature:
\begin{align}
    \braket{C_{\beta} | \hat{V} | C_{\beta^{\prime}}} &= \iiint_{\mathbb{R}^{3}} C_{\beta}(\vec{r})\, V(\vec{r})\, C_{\beta^{\prime}}(\vec{r})\, \mathrm{d}^{3}\vec{r} \nonumber\\
    &= \underbrace{\iiint_{\Omega_{m}} C_{\beta}(\vec{r})\, V(\vec{r})\, C_{\beta^{\prime}}(\vec{r})\, \mathrm{d}^{3}\vec{r}}_{\text{Becke grid}} + \underbrace{\iiint_{\Omega_{e}} C_{\beta}(\vec{r})\, V(\vec{r})\, C_{\beta^{\prime}}(\vec{r})\, \mathrm{d}^{3}\vec{r}}_{\text{FEDVR grid}}\ ,
\end{align}
where the left-hand side is evaluated with the Becke scheme described in this section, and the right-hand side is computed analogously to Eq.~(13). At the connector radius $R_m$, the integration weights must be partitioned between the inner and outer regions, which employ different quadrature rules. Because the connector does not follow the FEDVR quadrature rule, it should, in essence, be treated as belonging to the inner region.

\subsection{Construction of the orthonormal hybrid basis}
We now turn to the construction of an orthonormal hybrid MO–FEDVR basis in $\Omega_m$. The MOs already form an orthonormal set, $\langle\varphi_\sigma|\varphi_{\sigma^\prime}\rangle=\delta_{\sigma\sigma^\prime}$. By contrast, the FEDVR functions are only approximately orthonormal in $\Omega_m$ when evaluated with Becke’s quadrature. This limitation arises because Gauss-Lobatto quadrature is exact only for polynomials up to degrees $2n-3$, while the overlap between two FEDVRs is a polynomial of degree $2n-2$. In the external region, this does not compromise accuracy provided the Gauss–Lobatto rule is applied consistently and the FEDVR basis is sufficiently dense to represent the physical solution of the problem. To ensure full consistency between the FEDVRs and MOs in $\Omega_m$, and produce a robust orthonormalization procedure free from linear-dependence, we proceed in three steps:

\noindent\textbf{(1) Construct an othornomal basis of FEDVRs.} We consider the space of the $N_r$ primitive FEDVR functions 
$\{ \chi_1, \chi_2, \dots, \chi_{N_r} \}$ in $\Omega_m$ 
and diagonalize the associated overlap matrix
\begin{equation}
\hat{S} = \hat{U}\,\hat{\Lambda}\,\hat{U}^\dagger, 
\qquad 
S_{\mu\nu} = \langle \chi_\mu \mid \chi_{\nu} \rangle,
\end{equation}
where $\hat{\Lambda}$ is a diagonal matrix containing strictly positive eigenvalues 
and $\hat{U}$ is the unitary matrix of eigenvectors. 
The orthonormal FEDVR basis functions are then expressed as linear combinations
of the primitive ones,
\begin{equation}
\lvert \tilde{\chi}_\mu \rangle = \sum_{\nu=1}^{N_r} 
\lvert \chi_\nu \rangle \, C_{\nu\mu},
\end{equation}
with the orthogonalization matrix $\hat{C} = \hat{U} \hat{\Lambda}^{-1/2}$. By construction they satisfy 
$\langle \tilde{\chi}_\mu \mid \tilde{\chi}_\nu \rangle = \delta_{\mu\nu}$ 
under Becke’s quadrature rule.

\noindent\textbf{(2) Construct pure FEDVR functions orthogonal to the MOs.} We diagonalize the projector on the MO space
\begin{equation}
\hat{\mathcal{P}}_{0} = \sum_{\sigma=1}^{N_o} |\varphi_\sigma\rangle \langle \varphi_\sigma|,
\end{equation}
in the subspace spanned by the orthonormalized FEDVR functions $|\tilde{\chi}_{\mu}\rangle$, for $\mu=1,\dots,N_r$, yielding $N_r$ eigenvectors with eigenvalues $0\le \lambda_i \le 1$, for $i=1, \dots, N_r$, among which at least $N_r - N_o$ eigenvalues are zero. The corresponding eigenvectors $|\bar{\chi}^P_i\rangle$ ($\sigma=1,\dots,N_p$, with $N_p \ge N_r - N_o$) form, by construction, an orthonormal basis of pure FEDVRs orthogonal to the MO subspace.

\noindent\textbf{(3) Construct mixed MO–FEDVR functions orthogonal to the MOs.} 
To improve the spatial description, we refine the basis by considering the remaining eigenvectors 
$|\bar{\chi}^{(0)}_i\rangle$ with associated eigenvalues $\lambda_i>0$, for $i=N_p+1,\dots,N_r$. 
Since these eigenvectors have nonzero overlap with the MO space, we remove their MO components by constructing
\begin{equation}
|\bar{\chi}^{(1)}_i\rangle = (\hat{I}-\hat{\mathcal{P}}_0)\,|\bar{\chi}^{(0)}_i\rangle
= |\bar{\chi}^{(0)}_i\rangle - \sum_{\sigma=1}^{N_o}
\langle \varphi_\sigma \mid \bar{\chi}^{(0)}_i \rangle \, |\varphi_\sigma\rangle,
\qquad i=N_p+1,\dots,N_r,
\end{equation}
where $\hat{\mathcal{P}}_0$ is the MO projector. 
The functions $|\bar{\chi}^{(1)}_i\rangle$ are, by construction, orthogonal to all pure FEDVRs 
and to the MOs, but they do not yet form an orthonormal set. 
To orthonormalize them, we proceed as in step (1) by diagonalizing their overlap matrix,
\begin{equation}
\hat{S} = \hat{U}\,\hat{\Lambda}\,\hat{U}^\dagger, 
\qquad 
S_{ii'} = \langle \bar{\chi}^{(1)}_i \mid \bar{\chi}^{(1)}_{i'} \rangle,
\end{equation}
where $\hat{\Lambda} = \mathrm{diag}(\lambda_i)$. 
The mixing of FEDVR and MO functions can lead to near-linear dependencies in the hybrid basis; 
overcompleteness is signaled by very small eigenvalues in $\hat{\Lambda}$. 
We therefore discard eigenvectors with $\lambda_i < \epsilon$, where $\epsilon=10^{-8}$ is a chosen threshold. 
After this procedure, we retain $N_m \leq N_r - N_p$ (necessarily $N_m \le N_0$) additional mixed FEDVR–MO 
functions, denoted $|\hat{\chi}^M_i\rangle$ with $i=N_p+1,\dots,N_p+N_m$.

As a result, we obtain a fully orthonormal basis consisting of 
$N_0$ MOs, $N_p$ pure FEDVRs, and $N_m$ mixed MO–FEDVRs in $\Omega_m$, 
together with $N_e$ external FEDVRs in $\Omega_e$, forming a hybrid basis of $N_b$ functions. The bridge FEDVRs that connect $\Omega_m$ and $\Omega_e$ are included in $\Omega_m$ 
and are, by construction, also orthogonal to the external FEDVRs. To simplify notation, we denote the total set of $N_f = N_p + N_m + N_e$ FEDVR functions as 
$|\chi_\beta\rangle$, $\beta=1,\dots,N_f$, regardless of their type, 
bearing in mind that only the external FEDVRs satisfy the FEDVR quadrature rule, 
whereas the inner pure and mixed FEDVRs are treated as ordinary functions.

Finally, we note that the procedure outlined above does not take angular momentum or MO symmetry into account when performing the orthogonalization (though we note that the MOs themselves are symmetrized). Consequently, the FEDVR functions in $\Omega_m$ do not possess a well-defined symmetry. In contrast, the spherical scattering functions used to compute angularly resolved photoelectron observables do have a well-defined symmetry by construction. The method can be extended to enable the construction of symmetry-adapted FEDVRs, offering both improved efficiency and a clearer representation. However, since the number of FEDVRs in $\Omega_m$ is typically below $10^{4}$, even for the treatment of strong-field processes, the orthonormalization of the hybrid basis can in general be performed very rapidly. Using shared-memory parallel programming, this step takes only a few minutes on a personal computer.

\subsection{Time-Dependent Propagation}

The time-dependent Schr\"{o}dinger equation (TDSE) for an atom/molecule interacting with an electromagnetic field, using the dipole approximation and the length gauge, is given by
\begin{equation}
    i\partial_{t}\ket{\psi(t)} = \left[\hat{\mathcal{H}}_{0} + \hat{\mathcal{H}}_{\text{int}}(t)\right] \ket{\psi(t)}\ ,
    \label{eq:TDSE}
\end{equation}
where $\hat{\mathcal{H}}_{0}$ and $\hat{\mathcal{H}}_{\text{int}}(t) = \vec{r} \cdot \vec{E}(t)$ are, respectively, the field-free and interaction Hamiltonians, as described previously. The TDSE is propagated in the space spanned by $N$ hybrid basis functions, after discarding the FEDVR functions sitting at the end of the external grid.

To suppress spurious reflections of the outgoing wavepacket, we employ a complex absorbing potential (CAP), denoted $\hat{V}_{\text{CAP}}$, 
implemented as an imaginary quartic potential acting at the outermost region of the grid,
$V_{\text{CAP}}(r) = -i\,\eta \,(r-R_c)^4 \,\theta(r-R_c)$ ,
where $\eta>0$ controls the absorption strength, $R_c<R_e$ is the onset radius of the CAP, and $\theta$ is the Heaviside step function. 

The eigenstates $|\Phi_n\rangle$ of the system and their corresponding energies, defined by $\hat{\mathcal{H}}_0|\Phi_n\rangle = E_n|\Phi_n\rangle$, 
are obtained by diagonalizing $\hat{\mathcal{H}}_0$ with the unitary transformation $\hat{U}_0$, 
\begin{equation}
\hat{U}_0^\dagger \hat{\mathcal{H}}_0 \hat{U}_0 = \hat{E}, 
\qquad 
\hat{E} = \mathrm{diag}(E_0, E_1, \dots, E_{N-1}).
\end{equation}
Here, energies $E_n < 0$ correspond to electronic bound states, 
whereas energies $E_n > 0$ are associated with box-like states that describe the continuum dynamics. We choose the ground state $|\Phi_0\rangle$ as the initial state of the system, 
so that the initial condition in the TDSE [Eq.~(\ref{eq:TDSE})] is 
$|\psi(0)\rangle = |\Phi_0\rangle$.

To propagate the wavefunction in time, from a time $t$ to $t+\Delta t$, we employ the time-evolution operator $\hat{U}(t + \Delta t,t)$, such that $\ket{\psi(t + \Delta t)} = \hat{\mathcal{U}}(t + \Delta t, t) \ket{\psi(t)}$. The time-evolution operator $\hat{\mathcal{U}}(t + \Delta t,t)$ is approximated using a split-operator method:
\begin{align}\label{eq:spec_split}
    \hat{\mathcal{U}}(t + \Delta t, t) \approx e^{-i \hat{\mathcal{H}}_{0}\, \Delta t/2} e^{-i \hat{\mathcal{H}}_{\text{int}}(t + \Delta t/2)\, \Delta t} e^{-i \hat{\mathcal{H}}_{0}\, \Delta t/2} e^{-i \hat{V}_{\text{CAP}}\, \Delta t}\ .
\end{align}
As a local operator, $\hat{V}_{\text{CAP}}$ is diagonal in the basis of external FEDVRs and its action is thus trivial. The action of $\exp{\left[-i \hat{\mathcal{H}}_{\text{int}}(t)\Delta t\right]}$ is evaluated in the orthonormal hybrid basis, exploiting the sparsity of $\hat{H}_{\rm int}$ in the external region. Here, we evaluate the exponential using the matrix-inversion method (MIM) \cite{Alexei2010}.
On the other hand, the action of $\exp[-i\hat{\mathcal{H}}_{0}\,\Delta t/2]$ is carried out directly in the eigenbasis of the field-free Hamiltonian, after performing the unitary transformation: 
\begin{equation}
e^{-i\hat{\mathcal{H}}_{0}\,\Delta t} 
= \hat{U}_0 \, e^{-i \hat{E}\,\Delta t} \, \hat{U}_0^{\dagger}.
\end{equation}
This allows the field-free time propagation to be evaluated without approximating the exponential term $e^{-i\hat{H}_0\, \Delta t}$. In addition, since $\hat{\mathcal{H}}_{0}$ is time independent, its diagonal representation can be computed once before the time propagation and subsequently applied whenever needed.
 
In all calculations presented in this work, we employ a linearly polarized laser pulse with a sine-square envelope, given by
\begin{equation}\label{eq:sin2_field}
    \vec{E}(t) = E_{0} \sin^{2}\!\left(\frac{\pi t}{nT}\right) 
    \cos\!\left(\frac{2\pi t}{T}\right)\,\hat{\epsilon}, 
    \qquad 0 < t < nT,
\end{equation}
and zero otherwise. 
Here, $E_{0}$ is the field amplitude, $T$ the optical period, $\hat{\epsilon}$ the polarization direction, 
and $n$ the number of cycles.

\subsection{Observables}

\subsubsection{Photoelectron Spectra and Momentum Distributions}

Our procedure for computing photoelectron observables follows in many respects the one exposed in Ref.~\cite{borras2024}, with minor modifications in the application of the matching condition and the use of FEDVRs in place of B-splines. 

The energy-normalized spherical scattering states 
$\ket{\Psi^{-}_{\alpha E}}$ at fixed energy $E>0$, 
which fulfill incoming-wave boundary conditions, possessing a well-defined outgoing angular-momentum character $\alpha=(\ell_\alpha, m_{\ell_\alpha})$, have the asymptotic form
\begin{equation}
\Psi^{-}_{\alpha E}(\vec{r})
\xrightarrow[r\to\infty]{}
\sum_\beta \frac{u^{-}_{\beta,\alpha E}(r)}{r}\,
X_{\ell_\beta m_{\ell_\beta}}(\hat{r}),
\label{eq:scattering-incoming}
\end{equation}
where the scattering radial functions are given by
\begin{equation}
u^{-}_{\beta,\alpha E}(r)
= -i(2\pi k)^{-1/2}
\left[
\delta_{\alpha\beta}\,e^{i\Theta_\alpha(r)}
- S^{*}_{\beta\alpha}(E)\,e^{-i\Theta_\beta(r)}
\right].
\end{equation}
In the above equation, $\beta=(\ell_\beta,m_{\ell_\beta})$ is a collective index over the $M$ angular momentum channels, $\hat{S}$ is the scattering matrix, and the function
\begin{equation}
\Theta_\alpha(r) = kr + \frac{Z}{k}\ln(2kr) - \frac{\ell_\alpha\pi}{2} 
+ \sigma_{\ell_\alpha}(k),
\end{equation}
where $k=\sqrt{2E}$ is the electron momentum, $Z$ the effective charge of the asymptotic Coulomb potential acting on the photoelectron, 
and $\sigma_{\ell_\alpha}(k)=\arg[\Gamma(\ell_\alpha+1+iZ/k)]$ is
the Coulomb phase shift.

To construct spherical scattering states with the asymptotic behavior given in \eqref{eq:scattering-incoming}, we first build fixed-energy states $\ket{\Psi_{\alpha E}}$ from the box eigenstates $|\Phi_i\rangle$ of $\hat{\mathcal{H}}_0$. In this construction, the last FEDVR function  $|\chi_{N,n}^\alpha\rangle$, associated with angular channel $\alpha$, is reinstated at the boundary of the external grid ($R_e=R_{N,n}$). 
These states are then expressed as

\begin{equation}
    \ket{\Psi_{\alpha E}} = \ket{\chi_{N,n}^{\alpha}} 
    + \sum_{i=0}^{M} c^\alpha_{i} \ket{\Phi_{i}},
    \qquad 
    c^\alpha_{i} = \sum_{i^{\prime}=0}^{M} \frac{\braket{\Phi_{i^{\prime}} | \hat{\mathcal{H}}_0 | \chi_{N,n}^{\alpha}}}{E-E_{i^{\prime}}},
    \label{eq:scattering states}
\end{equation}
where $M$ is the total number of eigenstates. This can readily be shown by solving $\hat{\mathcal{H}}_0 \ket{\Psi_{\alpha E}} = E \ket{\Psi_{\alpha E}}$ and using the fact that $\hat{\mathcal{H}}_0|\Phi_{i}\rangle=E_{i}|\Phi_i\rangle$. Note that the above formula may fail if $E$ lies extremely close to one of the box eigenvalues $E_{n'}$, a case for which a prescription has been provided in Ref.~\cite{borras2024}. 
Such a situation is highly unlikely and therefore do not pose any practical issues. By construction, these scattering states $\ket{\Psi_{\alpha E}}$ vanish at the outer boundary of the grid for all angular components $\beta\ne\alpha$. At large distance, $r\to\infty$, each of these states behave as linear combination of regular $\mathcal{F}$ and irregular $\mathcal{G}$ Coulomb functions in each angular channel
\begin{equation}
\Psi_{\alpha E}(\vec{r})\xrightarrow[r\to\infty]{}\sum_\beta \frac{u_{\beta,\alpha E}(r)}{r}X_{\ell_\beta m_{\ell_\beta}}(\hat{r}),
\qquad 
u_{\beta,\alpha E}(r)=A_{\beta, \alpha E}\mathcal{F}_{\ell_\beta}(kr)+B_{\beta, \alpha E}\mathcal{G}_{\ell_\beta}(kr).
\end{equation}
The coefficients $A_{\beta, \alpha E}$ and $B_{\beta, \alpha E}$ can be obtained by matching the value of $u_{\beta,\alpha E}$ at the two last grid points to the known values of the regular and irregular functions at these points. This leads a system of linear equations for each $\alpha$ of the form 
\begin{equation}\label{eq:scat_sys}
    \begin{cases}
        A_{\beta, \alpha E}\, \mathcal{F}_{\ell_\beta}(kR_{N,n})+B_{\beta, \alpha E}\, \mathcal{G}_{\ell_\beta}(kR_{N.n}) = u_{\beta,\alpha E}(R_{N,n})\\
        A_{\beta,\alpha E}\, \mathcal{F}_{\ell_\beta}(kR_{N,n-1}) + B_{\beta,\alpha E}\, \mathcal{G}_{\ell_\beta}(kR_{N.n-1}) =u_{\beta,\alpha E}(R_{N,n-1})
    \end{cases}\ ,
\end{equation}
where the values of $u_{\beta,\alpha E}$ at the last two grid points are given by
\begin{equation}
 u_{\beta,\alpha E}(R_{N,n})=\frac{\delta_{\alpha\beta}}{\sqrt{w_{N,n}}},\quad u_{\beta,\alpha E}(R_{N,n-1})=\frac{1}{\sqrt{w_{N,n-1}}}\sum_{i}c_{i}^\alpha\langle \chi^\beta_{N,n-1}|\Phi_{i}\rangle.
\end{equation}
By solving the above system of equations, we can obtain the matrices $\mathbf{A}$ and $\mathbf{B}$, where $[\mathbf{A}]_{\beta,\alpha}=A_{\beta,\alpha E}$ and $[\mathbf{B}]_{\beta,\alpha}=B_{\beta,\alpha E}$. Knowing the asymptotic form of $\Psi_{\alpha E}$, it is now possible to use these functions as a basis to construct energy-normalized scattering states with incoming boundary conditions \eqref{eq:scattering-incoming}. Recalling that 
\begin{equation}
e^{\pm i\Theta_\alpha(r)}=\frac{\mathcal{G}_\alpha(r)\pm i\mathcal{F}_\alpha(r)}{2},
\end{equation}
one can show that the proper asymptotic form can be recovered through the following combination of our scattering basis functions:
\begin{equation}\label{eq:scat_states}
    \ket{\psi_{\alpha E}^{-}} =\sqrt{\frac{2}{\pi k}} \sum_{\beta} \ket{\psi_{\beta E}} \left[\frac{1}{\mathbf{A} + i \mathbf{B}}\right]_{\beta, \alpha},
\end{equation}
where the $\hat{\mathcal{S}}$ matrix takes the form:
\begin{equation}\label{eq:s-matrix}
\hat{\mathcal{S}}=\frac{\mathbf{A}+i\mathbf{B}}{\mathbf{A}-i\mathbf{B}}.
\end{equation}

From these scattering states, we can compute the total photoionization cross section  
\begin{equation}\label{eq:PCS}
    \sigma(E) = \frac{4\pi^{2} \omega}{3c} \sum_{\alpha} \big{|} \braket{\psi_{\alpha E}^{-} | \vec{r} | \Phi_{0}}\big{|}^{2},
\end{equation}
where $\omega=E-E_0$ is the light frequency to ionize an electron with asymptotic energy $E$ from an initial state $\ket{\Phi_{0}}$ with energy $E_0$. We can also find the energy-resolved ionization probability by projecting onto the final wavefunction $\ket{\psi(t_f)}$ at the end of the pulse, $t_f$, leading to:
\begin{equation}\label{eq:PES}
    \frac{dP}{dE} = \sum_{\alpha} \big{|} \braket{\psi^{-}_{\alpha E} | \psi(t_f)} \big{|}^{2}.
\end{equation}

Finally, it is possible to construct a scattering wave function $|\psi^-_{E\hat{k}}\rangle$ for a photoelectron with a well-defined asymptotic momentum $\vec{k}$ using the spherical scattering functions, namely
\begin{equation}
|\psi^-_{E\hat{k}}\rangle=\sum_{\alpha}i^{\ell_\alpha}e^{-i\sigma_{l_\alpha}(k)}X_{\ell_\alpha m_{\ell_\alpha}}(\hat{k})\ket{\psi_{\alpha E}^{-}}.
\end{equation}
As a result, the photoelectron momentum distribution (PMD) for an electron emitted in a direction $\hat{k}$ with energy $E=k^2/2$ is given by
\begin{equation}\label{eq:PMD}
    \frac{\mathrm{d}P}{\mathrm{d}E\,\mathrm{d}\Omega_{\hat{k}}} = \bigg{|} \sum_{\alpha} i^{-\ell_\alpha}\, X_{\ell_\alpha m_{\ell_\alpha}}(\hat{k})\, e^{i \sigma_{\ell_\alpha}(k)} \braket{\psi^{-}_{\alpha E} | \psi(t_f)} \bigg{|}^{2}.
\end{equation}

Although the procedure outlined above for computing photoelectron observables was presented for single-electron systems, it generalizes straightforwardly to multielectron targets, including multiple ionization channels, in the same spirit as given in Ref.~\cite{borras2024}.

\subsubsection{High-Harmonic Generation Spectra}

High-harmonic spectra are computed using the time-dependent dipole moment of the system, defined by $ \vec{\mu}(t) = \braket{\psi(t) | \vec{r}| \psi(t)}$. The Fourier transform of the dipole response is then given by
\begin{equation}
    \tilde{\mu}_i(\omega) =  \mathcal{F}\left[W(t) \cdot
    \mu_{i}(t)\right] 
\end{equation}
where $\mu_{i}(t) = \vec{\mu}(t) \cdot \hat{r}_{i}$, $i=x,y,$ and $z$ are the components of the dipole moment in each of the three Cartesian directions and $\tilde{\mu}_i$ their corresponding Fourier transform. The time-dependent dipole moment $\vec{\mu}(t)$ is calculated for one full optical cycle after the laser pulse is finished, and the window function $W(t)$ in the above equation is a function which is 1 until the end of the laser pulse and smoothly goes to 0 over the next optical cycle. This window function is intended to suppress any non-physical high-frequency oscillations due to the non-zero value of the dipole moment (and its derivative) at the end of the calculation. The optical density is then expressed as
\begin{equation}\label{eq:HHG_spec}
S(\omega)=\frac{2\omega^4}{3\pi c^3}\sum_{i=1}^3\left|\tilde{\mu}_i(\omega)\right|^2
\end{equation}
For a spherically-symmetric system such as atomic hydrogen, the only nonzero component of $\vec{\mu}(t)$ is parallel to the axis of the laser polarization, but this is not necessarily the case for more complex systems.


\section{Results and Discussion}

\subsection{Atomic Hydrogen (H) and One-electron Model in Helium (He)}

We describe atomic hydrogen using a hybrid basis of GTOs and FEDVR functions. The GTO set comprises three \(s\)-type primitives with exponents \(5.085\), \(2.0\), and \(0.30\), and three \(p\)-type primitives with exponents \(3.085\), \(1.0\), and \(0.50\). This small GTO basis set alone is by itself incapable of giving a good description of any of the bound states of atomic hydrogen. The GTO basis is augmented with FEDVRs, and the hydrogen energies are then obtained by diagonalizing \(\hat{\mathcal{H}}_{0}\) in this orthonormalized hybrid basis. Table~\ref{tab:1} lists the energies of the \(ns\) states up to principal quantum number \(n=9\), obtained with \(R_m=30~\mathrm{a.u.}\) and \(R_e=350~\mathrm{a.u.}\). The relative error of the \(1s\) energy is on the order of \(10^{-13}\). As \(R_e\) is increased, the number of converged \(ns\) states grows, consistent with their characteristic radius scaling as \(n^2\). For \(R_e=350~\mathrm{a.u.}\), the results reported in Table~\ref{tab:1} for the first nine \(s\)-type states show absolute errors near machine precision (\(\sim 10^{-15}\)) and relative errors no larger than \(10^{-11}\) for any \(ns\) state. Importantly, the results are highly stable with respect to the inner-grid radius \(R_m\): they remain unchanged as \(R_m\) is varied from \(20~\mathrm{a.u.}\) (beyond which all GTOs are vanishingly small) to larger values. The results are also stable upon increasing the density of FEDVR functions further, including cases where some hybrid functions are discarded to avoid linear dependencies. This robustness validates the procedures used to construct the orthonormal basis and to evaluate the one-electron integrals entering the field-free Hamiltonian.

\setlength{\tabcolsep}{4pt}
\renewcommand{\arraystretch}{1.25}
\begin{table}[htbp]
    \centering
    \begin{tabular}{|c|c|c|c|}
        \hline
        $ns$ & Exact Energy & Abs. Error & Rel. Error\\
        \hline \hline
        $1s$ & $-\sfrac{1}{2}$ & $7.2 \times 10^{-14}$ & $1.4 \times 10^{-13}$\\
        \hline 
        $2s$ & $-\sfrac{1}{8}$ & $3.5 \times 10^{-14}$ & $2.8 \times 10^{-13}$\\
        \hline 
        $3s$ & $-\sfrac{1}{18}$ & $3.0 \times 10^{-13}$ & $5.3 \times 10^{-12}$\\
        \hline 
        $4s$ & $-\sfrac{1}{32}$ & $2.1 \times 10^{-13}$ & $6.7 \times 10^{-12}$\\
        \hline 
        $5s$ & $-\sfrac{1}{50}$ & $5.4 \times 10^{-15}$ & $2.7 \times 10^{-13}$\\
        \hline 
        $6s$ & $-\sfrac{1}{72}$ & $1.1 \times 10^{-14}$ & $8.2 \times 10^{-13}$\\
        \hline 
        $7s$ & $-\sfrac{1}{98}$ & $1.2 \times 10^{-14}$ & $1.2 \times 10^{-12}$\\
        \hline 
        $8s$ & $-\sfrac{1}{128}$ & $4.4 \times 10^{-15}$ & $5.6 \times 10^{-13}$\\
        \hline 
        $9s$ & $-\sfrac{1}{162}$ & $1.1 \times 10^{-14}$ & $1.8 \times 10^{-12}$\\
        \hline
    \end{tabular}
    \caption{Exact energies of the $ns$ states of hydrogen, and the absolute and relative errors of the energies calculated with ATTOMESA. All quantities are in atomic units.}
    \label{tab:1}
\end{table}
We performed several benchmarks to validate our description of atomic hydrogen in an external electric field. First, we estimated the static polarizability $\alpha \equiv \mu_{z} / E_{z}$ using a long-wavelength (3-$\mu$m), low-intensity ($3.5 \times 10^{8}\, \text{W}/\text{cm}^{2}$), 3-optical-cycle laser pulse. Because the laser field is very weak, we only considered $s$- and $p$-type states. With ATTOMESA, we obtained a value of $\alpha = 4.506$ a.u., to be compared with the exact value of 4.5 atomic units.  The small difference in these two values is mainly attributed to the slight non-adiabaticity of the laser field. 

\begin{figure}[h]
    \centering
    \includegraphics[width=0.95\columnwidth]{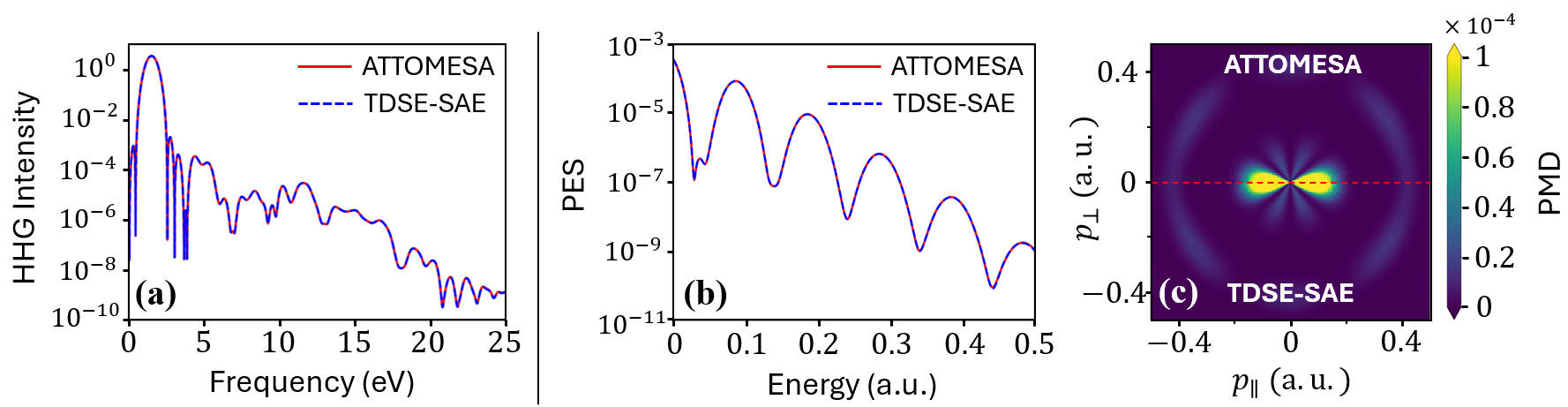}
    \caption{Comparisons between ATTOMESA and TDSE-SAE in atomic hydrogen: (a) HHG with an 800-nm, $2.2 \times 10^{13}\ \text{W}/\text{cm}^{2}$, 3-o.c. laser pulse; (b) photoelectron spectrum and (c) photoelectron momentum distribution for a $2.72$-eV, $7.5\times 10^{11}\ \text{W}/\text{cm}^{2}$, 10-o.c. laser pulse.}
    \label{fig:2}
\end{figure}

Second, we compared the HHG spectrum obtained with \textsc{ATTOMESA} to that from our implementation of the time-dependent Schr\"odinger equation in the single-active-electron approximation (TDSE–SAE) \cite{douguet2016, douguet2024}, which is extremely accurate for atomic hydrogen. The HHG spectra calculated with both codes are shown in Figure~{\ref{fig:2}}(a). Here, we use an 800-nm, $2.2 \times 10^{13}\, \text{W}/\text{cm}^{2}$, 3-optical-cycle laser pulse for both calculations, including up to $\ell \leq 10$ angular momentum which is sufficient to describe this low-intensity process. The agreement between the two methods is again excellent. 

Lastly, extending the outer grid to 500 atomic units, we computed the photoelectron spectrum under the influence of a 2.72-eV, $7.5 \times 10^{11}\ \text{W}/\text{cm}^{2}$, 10-optical-cycle laser pulse. In Figure~{\ref{fig:2}(b)}, we compare the photoelectron spectrum calculated with Eq.~(\ref{eq:PES}), and in Fig.~{\ref{fig:2}}(c), we compare the photoelectron momentum distributions using Eq.~(\ref{eq:PMD}). Again, we consider only states with $\ell \leq 10$ and in the calculations and found excellent agreement between ATTOMESA and our TDSE-SAE code.

\begin{figure}[h]
    \centering
    \includegraphics[width=0.95\columnwidth]{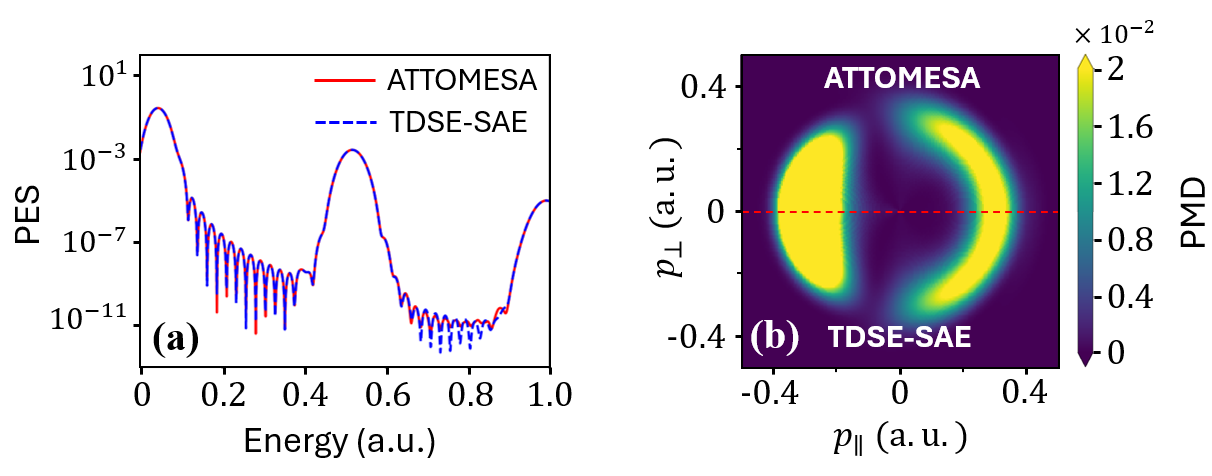}
    \caption{Comparison between ATTOMESA and TDSE-SAE in helium, using the Tong-Lin potential \cite{tong2005}: (a) photoelectron spectrum and (b) photoelectron momentum distribution using a $\omega+2\omega$ scheme ($\omega = 0.475$ a.u., with an intensity of $8.8 \times 10^{13}$ W/cm$^{2}$; $2\omega = 0.95$ a.u., with an intensity of $3.5 \times 10^{12}$ W/cm$^{2}$). Both pulses have the same duration ($6.4$ fs) and CEP ($-\pi/2$).}
    \label{fig:3}
\end{figure}

We also applied the procedure outlined in Sec.~2.7.1 to compute photoelectron observables in atoms described by SAE model potentials. Because these potentials coincide with the Coulomb tail at long range but differ at short range, the scattering states acquire, in addition to the Coulomb phase, an extra scattering phase shift \(\delta_{\ell}(k)\) directly related to the elements of the scattering matrix $S_\ell=e^{2i\delta_\ell(k)}$. This makes them a stringent benchmark for PMDs, which are highly phase-sensitive. In particular, we employed the Tong-Lin helium potential~\cite{tong2005} and drove ionization with a bichromatic \(\omega+2\omega\) scheme, so that one- and two-photon pathways interfere in the angular distribution, thus providing a direct test of the approach despite the spherical symmetry of the electrostatic potential. Using \(\omega=0.475\) a.u.\ and setting the intensity to obtain comparable signals from the one- and two-photon pathways, we obtained the total ionization probability and the PMDs shown in Fig.~\ref{fig:3}.

\subsection{Dihydrogen Cation (H$_{2}^{+}$)}

To validate the method when using more than one atom, we apply the hybrid basis to \(\mathrm{H}_2^+\) at an internuclear distance \(R_0=2~\mathrm{a.u.}\), replicating a benchmark similar to the one of Rescigno \emph{et al.}~\cite{rescigno2005}. The system is described using two atomic shells centered on the nuclei with radii of \(30~\mathrm{a.u.}\), while the master shell extends to $R_m=40~\mathrm{a.u.}$ We employ the same Gaussian basis than in \cite{rescigno2005}, taking the six functions with the largest exponents (exponents 1.20, 3.38, 10.60, 38.65, 173.58, and 1170.50) from Huzinaga’s $10s$ expansion of the hydrogen $1s$ function in Gaussians \cite{huzinaga1965}, as well as four $p$-type Gaussians (exponents 0.325, 0.75, 1.50, and 3.00). Again we note that this basis set alone provides a very poor description of all bound states of H$_2^+$.

We supplement the GTO basis with FEDVR functions defined on a radial grid centered at the molecular midpoint and extending to \(R_{e}=100~\mathrm{a.u.}\). As a first test, we examine the convergence of the ground electronic state \(1s\,\sigma_{g}\) of \(\mathrm{H}_{2}^{+}\) with respect to the number of angular–momentum channels associated with the FEDVRs (Table~\ref{tab:2}). Using only the Gaussian basis, we obtain an energy of \(-0.8329\)~a.u., which deviates markedly from the accuate value \(-1.1026\)~a.u. reported by Madsen and Peek~\cite{madsen1970}. Upon adding FEDVR basis functions with \((\ell,m)=(0,0)\) only, the energy already reproduces the first four significant digits of the \(1s\,\sigma_{g}\) benchmark. Increasing \(\ell_{\max}\) further systematically improves the result, reaching agreement at the level of six significant figures for \(\ell_{\max}\!\geq\!10\), beyond which no further improvement is observed with the radial FEDVR grid chosen.

\begin{table}[h]
    \centering
    \begin{tabular}{|c|c|c|}
        \cline{2-3}
        \multicolumn{1}{c|}{} & Energy (a.u.) & Rel. Error\\
        \hline \hline
        Gaussians only & -0.8329533 & $2.45 \times 10^{-1}$\\
        \hline
        $\ell_{\text{max}} = 0$ & -1.102(3766) & $2.34 \times 10^{-4}$\\
        $\ell_{\text{max}} = 2$ & -1.1026(282) & $5.44 \times 10^{-6}$\\
        $\ell_{\text{max}} = 6$ & -1.10263(22) & $1.81 \times 10^{-6}$\\
        $\ell_{\text{max}} = 10$ & -1.10263(28) & $1.25 \times 10^{-6}$\\
        $\ell_{\text{max}} = 14$ & -1.10263(27) & $1.36 \times 10^{-6}$\\
        \hline
        Reference \cite{madsen1970} & -1.1026342 & 0\\
        \hline
    \end{tabular}
    \caption{Convergence of the $1s \sigma_{g}$ ground state of H$_{2}^{+}$ with increasing number of angular momenta at its equilibrium internuclear distance.}
    \label{tab:2}
\end{table}

The principal limiting factor of the hybrid-basis accuracy, especially for the \(\mathrm{H}_2^+\) ground state, is the difficulty of representing the electron–nuclear cusp at off-center nuclei. We find that refining the FEDVR mesh in the nuclear vicinity, together with increasing \(\ell_{\max}\), improves the energy, albeit slowly. Moreover, augmenting the Huzinaga GTO set with a very tight primitive (exponent \(\alpha=10^{4}\)) sharpens the cusp description and yields nearly seven significant figures of accuracy.

In Table~\ref{tab:3}, we report the energies of the lowest fourteen electronic states of \(\mathrm{H}_2^+\) at \(R_0=2~\mathrm{a.u.}\) computed with \textsc{ATTOMESA}, alongside the reference values of Madsen and Peek~\cite{madsen1970}. We observe  an overall excellent agreement for all of the excited states.

\begin{table}[h]
    \centering
    \begin{tabular}{|c|c|c|c|}
        \hline
        State & Reference \cite{madsen1970} & ATTOMESA & Rel. Error\\
        \hline \hline
        $1s\sigma_{g}$ & -1.10263421 & -1.10263(282) & $1.3 \times 10^{-6}$\\
        $2p\sigma_{u}$ & -0.66753439 & -0.66753(319) & $1.8 \times 10^{-6}$\\
        $2p\pi_{u}$ & -0.42877182 & -0.428771(03) & $1.8 \times 10^{-6}$\\
        $2s\sigma_{g}$ & -0.36086488 & -0.360864(52) & $1.0 \times 10^{-6}$\\
        $3p\sigma_{u}$ & -0.25541317 & -0.255412(98) & $7.1 \times 10^{-7}$\\
        $3d\sigma_{g}$ & -0.23577763 & -0.235777(53) & $4.3 \times 10^{-7}$\\
        $3d\pi_{g}$ & -0.22669963 & -0.226699(44) & $8.3 \times 10^{-7}$\\
        $3p\pi_{u}$ & -0.20086483 & -0.200864(54) & $1.4 \times 10^{-6}$\\
        $3s\sigma_{g}$ & -0.17768105 & -0.177680(91) & $7.3 \times 10^{-7}$\\
        $4p\sigma_{u}$ & -0.13731292 & -0.1373128(9) & $2.5 \times 10^{-7}$\\
        $4d\sigma_{g}$ & -0.13079188 & -0.1307918(7) & $4.7 \times 10^{-7}$\\
        $4d\pi_{g}$ & -0.12671013 & -0.126710(03) & $8.1 \times 10^{-7}$\\
        $4f\sigma_{u}$ & -0.12664387 & -0.126643(93) & $4.7 \times 10^{-7}$\\
        $4f\pi_{u}$ & -0.12619890 & -0.1261989(3) & $2.2 \times 10^{-7}$\\
        \hline
    \end{tabular}
    \caption{Energies for the first 14 states of H$_{2}^{+}$. All energies are listed in Hartrees.}
    \label{tab:3}
\end{table}

Next, we benchmarked the \textsc{ATTOMESA} static dipole polarizability of \(\mathrm{H}_2^+\) in the low-frequency, weak-field limit against the calculations of Rahman~\cite{rahman1953}. We obtained \(\alpha_{\parallel}=5.08\) and \(\alpha_{\perp}=1.76\)~a.u., in close agreement with the reference values \(\alpha_{\parallel}=5.06\) and \(\alpha_{\perp}=1.75\)~a.u. Residual discrepancies may be attributed to (i) slight nonadiabaticity of the applied field, (ii) the density of the FEDVR mesh, and (iii) the number of angular-momentum channels.

\begin{figure}[h]
    \centering
    \includegraphics[width=0.95\columnwidth]{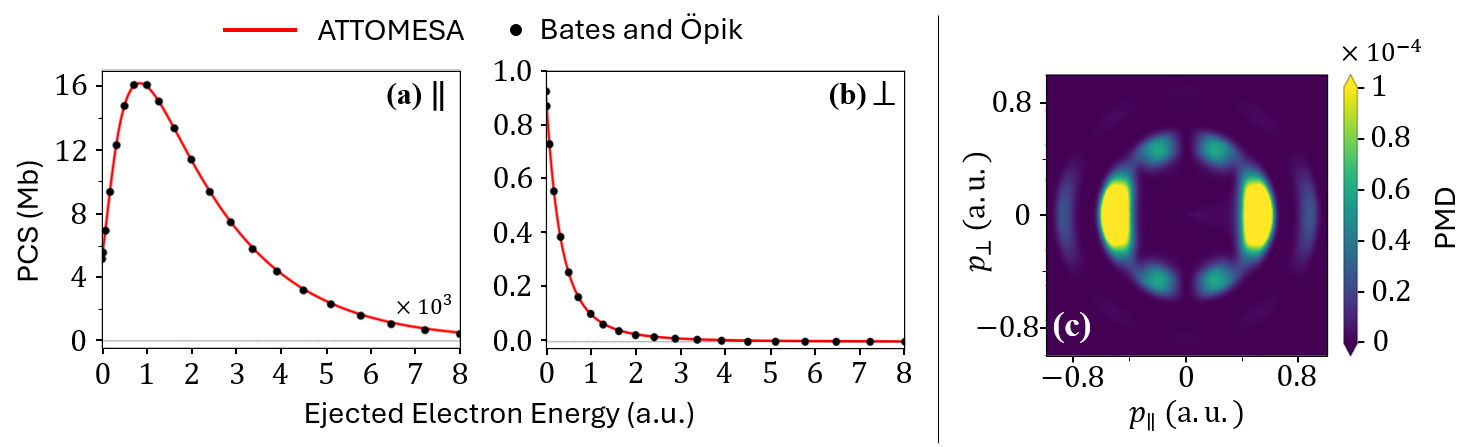}
    \caption{Total photoionization cross section for H$_{2}^{+}$, polarized (a) parallel to and (b) perpendicular to the molecular axis: the blue curve is from ATTOMESA, while the black dots are from Bates and \"{O}pik \cite{bates1968}; (c) photoelectron momentum distribution for H$_{2}^{+}$ using a 6.8-eV, $1.2 \times 10^{12}\ \text{W}/\text{cm}^{2}$, 10-o.c. laser pulse.}
    \label{fig:4}
\end{figure}

We also computed the one-photon total photoionization cross sections using Eq.~(\ref{eq:PCS}) for light polarized parallel and perpendicular to the molecular axis, shown in Figs.~\ref{fig:4}(a) and \ref{fig:4}(b), respectively. For these calculations we set the external radius to \(R_{e}=50~\mathrm{a.u.}\) and substantially refined the FEDVR mesh to better resolve the high-energy portion of the PCS, up to \(8~\mathrm{a.u.}\) (\(\approx 217.7~\mathrm{eV}\)). The red curve reports the \textsc{ATTOMESA} results, which are in excellent agreement with the reference numerical values of Bates and \"{O}pik~\cite{bates1968} (black dots).

Finally, we employed a 6.8-eV, $1.2 \times 10^{12}\ \text{W}/\text{cm}^{2}$, 10-o.c.\ laser pulse polarized along the molecular axis to ionize ground-state H$_2^+$ via a five-photon process. The resulting PMD is shown in Fig.~\ref{fig:4}(c).


\section{Conclusions and Perspectives}

In this work, we have presented the first comprehensive validation of the ATTOMESA code, focusing on benchmarks of one-electron dynamics. Specifically, we tested the construction of the hybrid Gaussian–FEDVR basis and quadrature scheme, the orthonormalization procedure, the time-propagation algorithm, and the computation of physical observables. These benchmarks establish the accuracy and robustness of the approach for single-electron systems, and, more importantly, provide the foundational framework that enabled a seamless and efficient integration with quantum-chemistry methods for treating correlated multi-electron dynamics.  

Beyond this technical demonstration, this article is also dedicated to the memory of our dear colleague and friend Barry Schneider. His pioneering work and vision were instrumental in shaping many of the ideas implemented here, and his generosity, insight, and encouragement continue to inspire this effort.

Building on this foundation, ATTOMESA has now been extended and integrated with quantum-chemistry methods for multi-electron systems, using numerical tools closely related to those employed in the complex Kohn variational method developed at Berkeley \cite{rescigno1995}. A detailed description of this extension will be presented in a forthcoming communication. Together, these advances establish ATTOMESA as a flexible and accurate framework for simulating time-dependent processes in correlated electronic systems, while carrying forward Barry’s legacy in computational attosecond science.

\vspace{6pt} 
\textbf{Author Contributions:} Conceptualization, K.A.H., N.D., L.A., and H.G.; methodology, N.D. and L.A.; software, K.A.H., N.D., and H.G..; validation, K.A.H.; formal analysis, K.A.H.; investigation, K.A.H. and N.D.; resources, N.D.; data curation, K.A.H.; writing---original draft preparation, K.A.H. and N.D.; writing---review and editing, K.A.H., N.D., L.A., and H.G.; visualization, K.A.H.; supervision, N.D.; project administration, N.D.; funding acquisition, N.D.. All authors have read and agreed to the published version of the manuscript.

\textbf{Funding:} N.D. acknowledges support from the National Science Foundation under Grant No. OAC-2311928. L. A. acknowledges support from the National Science Foundation under grant No. 2309133.

\textbf{Conflicts of Interest:} The authors declare no conflicts of interest.

\bibliographystyle{apsrev}
\bibliography{bibliography.bib}
\end{document}